\documentclass[multphys,vecphys]{svmult}

\usepackage{makeidx}   
\usepackage{graphicx}  
\usepackage{multicol}  

\usepackage{cite}

\makeindex             

\begin{document}

\title{Invasive Allele Spread under Preemptive Competition}

\titlerunning{Invasive Allele Spread under Preemptive Competition}

\author{J.A. Yasi\inst{1}, G. Korniss\inst{1} \and T. Caraco\inst{2}}

\authorrunning{Yasi et al.}

\institute{
Department of Physics, Applied Physics, and Astronomy, \\
Rensselaer Polytechnic Institute, 110 8th Street, Troy, NY
12180-3590, USA \\
\texttt{E-mail: yasij@rpi.edu, korniss@rpi.edu}
\and
Department of Biological Sciences, \\
University at Albany, Albany NY 12222, USA \\
\texttt{E-mail: caraco@albany.edu}
}

\maketitle

\begin{abstract}
We study a discrete spatial model for invasive allele spread in
which two alleles compete preemptively, initially only the ``residents"
(weaker competitors) being present. We find that the spread of the
advantageous mutation is well described by homogeneous nucleation; in particular,
in large systems the time-dependent global density of the resident allele is
well approximated by Avrami's law.
\end{abstract}

\section{Introduction and Model}

The spatial and temporal characteristics of the spread of an
advantageous mutation are fundamental questions in population
dynamics. Fisher \cite{FISHER_37} and Kolmogorov et al.
\cite{KOLMOGOROV_37} first addressed these questions
using the framework of a simple reaction-diffusion equation
\cite{MURRAY_03}. That work and many others focused on the
velocity of the propagating front, which exists initially and
separates the two spatial regions occupied separately by the two alleles.
Both continuum and discrete spatial models have successfully
tackled various aspects of these problems \cite{MURRAY_03,AVRAHAM_00}.
In this work we investigate how the advantageous allele emerges
from ``scratch"; i.e., initially the region is fully dominated by
the resident allele and the advantageous allele is introduced by
rare mutations. While the mutant allele has an individual-level
advantage over the original one, the low probability of mutations,
combined with a discrete spatial dynamics, can prevent the spread
of the mutant for long times. Here we consider a model where the
original ``resident" and the competitively superior ``invasive"
allele compete for a common limiting resource preemptively
\cite{TANEY_00,YU_01,AMARA_rev_03,SHURIN_04}.

The details of our model are as follows. We consider an
$L$$\times$$L$ lattice with periodic boundary conditions. Each
site can be empty or occupied by a {\em single} allele (either a
resident or an invader). A lattice site represents the minimal level of  locally
available resource required to sustain an individual organism,
hence the ``excluded volume" constraint. We introduce the
local occupation numbers at site ${\bf x}$, $n_{i}({\bf x})=0,1$,
$i=1,2$, representing the number of resident and invader alleles,
respectively. By virtue of the excluded volume constraint,
$n_{1}({\bf x})n_{2}({\bf x})=0$. New individuals arise
through local clonal propagation only. That is, an individual occupying
site ${\bf x}$ may reproduce if one or more neighboring sites are
empty (here we consider nearest neighbor colonization only).
Competition for resources, hence space, is preemptive, therefore,
an occupied site cannot be colonized by either allele until the
current occupant's mortality leaves that site empty. Each
individual may mutate and so carry the alternate allele;
mutation is a two-way, recurrent process.

We performed dynamic Monte Carlo simulations to study the above
model. Our time unit is one Monte Carlo step per site (MCSS)
during which $L^2$ sites are chosen randomly. If a site is empty,
it may be colonized by individuals of allele $i$ occupying
neighboring sites, at the rate $\alpha_i\eta_i({\bf x})$;
$\alpha_i$ is the individual-level colonization rate and
$\eta_i({\bf x})=(1/4)\sum_{{\bf x'}\epsilon {\rm nn}({\bf
x})}n_{i}({\bf x'})$ is the density of allele $i$ around site
${\bf x}$ [${\rm nn}({\bf x})$ is the set of nearest neighbors of
site $\bf x$]. If a site is occupied by an individual, it can die
at rate $\mu$ (regardless of the allele) or mutate to the other
allele at rate $\phi$. We can summarize the local transition rules
for an arbitrary site ${\bf x}$ as
\begin{equation}
0\stackrel{\alpha_1\eta_1(\bf{x})}{\longrightarrow}1, \;\;
0\stackrel{\alpha_2\eta_2(\bf{x})}{\longrightarrow}2, \;\;
1\stackrel{\mu}{\longrightarrow}0, \;\;
2\stackrel{\mu}{\longrightarrow}0, \;\;
1\stackrel{\phi}{\longleftrightarrow}2, \;
\label{rates}
\end{equation}
where $0,1,2$ indicates whether the site is empty, occupied by an
individual with the resident, or an individual with the invader
allele, respectively.

In the simulations, we initialized the system
fully occupied by the resident allele ($n_{1}({\bf x})$$=$$1$ for
all ${\bf x}$). We are interested in the parameter region where
$\phi\ll\mu<\alpha_1<\alpha_2$, so that mutation is a rare
process, but the invader allele has a reproductive-effort
advantage. Then, due to mortality, the system quickly relaxes
(much too fast for mutation to play a role) to the
``quasi-equilibrium" state where the resident's population is
balanced by its own clonal propagation and mortality rates
(in the near absence of invaders). Throughout the
simulations, we track the time-dependent global densities of the two
alleles, $\rho_i(t)=(1/L^2)\sum_{\bf x}n_{i}({\bf x},t)$. We
define the lifetime $\tau$ of the resident allele as the first
passage time of $\rho_1(t)$ to one-half of its quasi-equilibrium
value $\rho_{1}^{*}$.

\section{Single-Cluster and Multi-Cluster Spread}

As a result of the rare mutations, individuals with the invasive
allele occasionally appear in the population. An invader lacking access to nearby resources may die
without propagating. If a site
opens in the local neighborhood (resource becomes available),
the invader may colonize it. However, the empty
site is likely surrounded by more than one resident. The
resident's greater local density can compensate for its lower
individual-level colonization rate, so the resident has the better
chance of colonizing an empty site. Consequently, one expects small
clusters of the invading allele to shrink and disappear. Residents,
although weaker competitors, can prevail for some time, since
preemptive competition imposes a strong constraint on the growth
of the invaders. Individuals with the advantageous allele can
succeed only if they generate a cluster large enough that it
statistically tends to grow at its periphery.

Snapshots of configurations and preliminary studies \cite{OAKC_SPIE}
confirm the existence of a critical
cluster size, beyond which the spread of the invading allele
becomes statistically favorable. Further, they also
show strongly clustered growth of the invading allele. For a
given set of parameters, there exists a length scale $R_o$, the
typical spatial separation of invading clusters; for $L$$\ll$$R_o$
the invasion almost always occurs through the spread of a single
invading cluster [single-cluster (SC) invasion], while for
$L$$\gg$$R_o$ the invasion is the result of many invading clusters
[multi-cluster (MC) invasion]. Conversely, fixing the linear
system size $L$ and other parameters (except the mutation rate)
there is a characteristic value of $\phi$ (now controlling $R_o$), such
that for sufficiently low values of $\phi$, MC invasion by the
advantageous allele crosses over to the SC pattern.
\begin{figure}[t]
\centering
\includegraphics[width=.48\textwidth,height=4.7truecm]{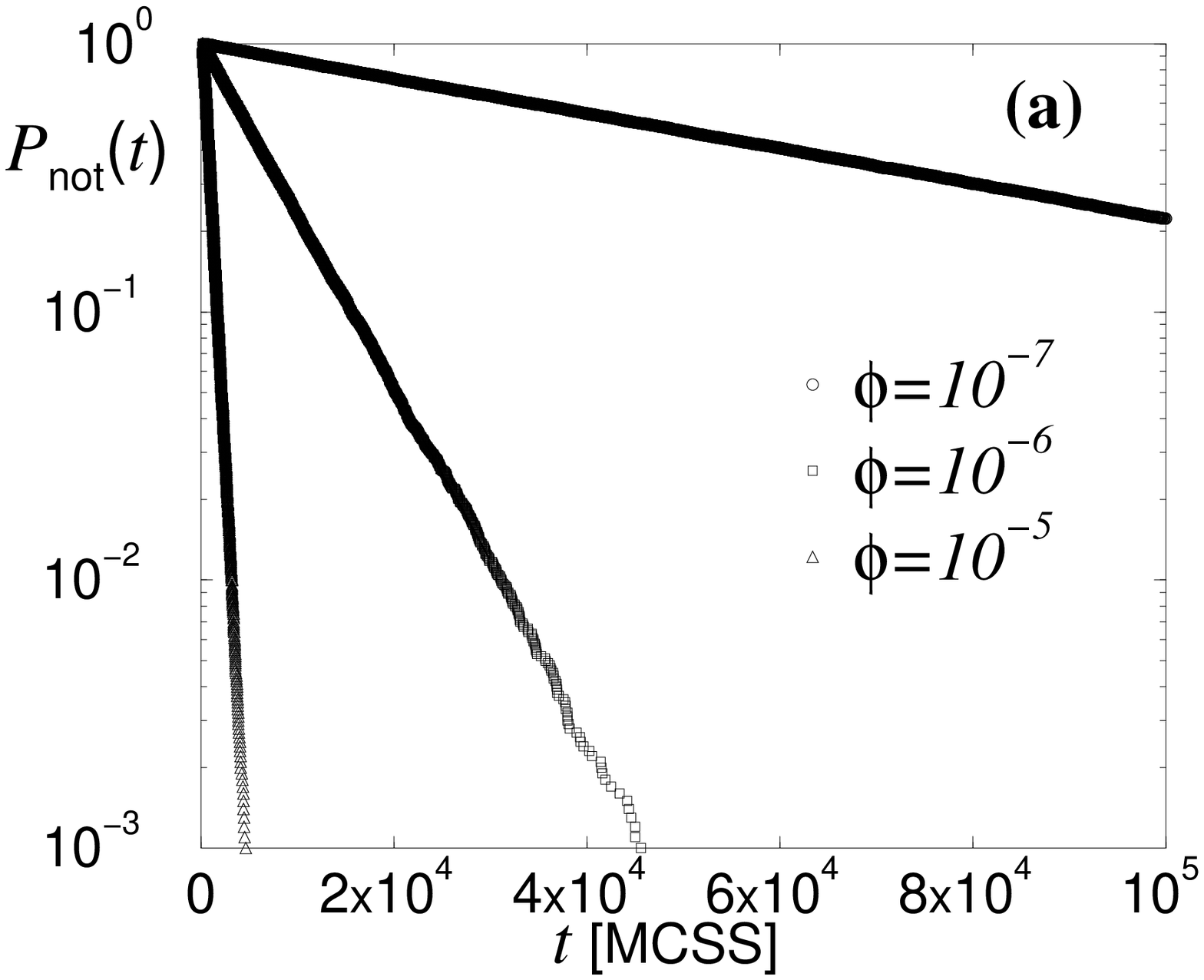}
\hspace*{0.2truecm}
\includegraphics[width=.48\textwidth,height=4.7truecm]{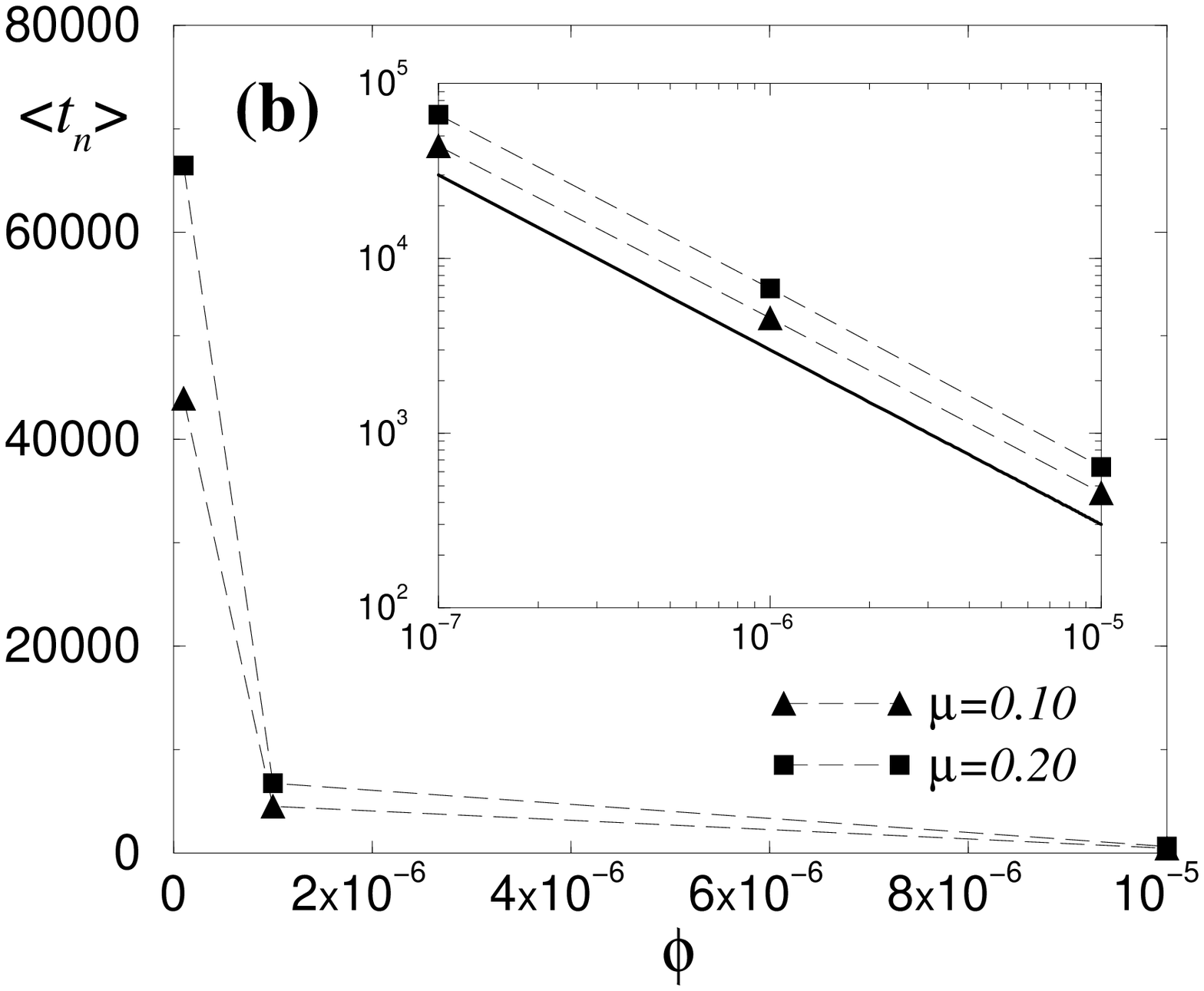}
\caption[]{
({\bf a}) Cumulative probability distributions $P_{\rm
not}(t)$ for $L$$=$$32$, $\alpha_1$$=$$0.50$, $\alpha_2$$=$$0.70$,
and $\mu$$=$$0.20$ for three different values of the mutation rate
$\phi$ (in increasing order from the top).
({\bf b}) Average nucleation time (in units of MCSS) vs. the mutation rate
for two different values of $\mu$ [$\alpha_1$, $\alpha_2$, and $L$
are the same as for (a)]. The inset shows the same on log-log
scales. The straight solid line corresponds to a slope $-1$,
indicating $\langle t_n\rangle\sim\phi^{-1}$ in the single-cluster
regime.}
\label{fig1}
\end{figure}

The above picture suggests that we can apply the framework of
homogeneous nucleation and growth \cite{KOLMOGOROV,JM,AVRAMI} to
describe the spatial and temporal characteristics of the spread of
the invasive allele. This framework has successfully described
analogous dynamic phenomena in
ferromagnetic \cite{RIKVOLD_94,RICHARDS_95,RAMOS_99}
and ferroelectric materials \cite{ISHIBASHI_71,DUIKER_90},
flame propagation in slow combustion \cite{KART98},
chemical reactions \cite{MACHADO_05},
and other ecological systems \cite{OAKC_SPIE,KC_JTB05,GANDHI_JTB99}.
While local mutation is a Poisson
process, lacking a Hamiltonian or an effective free energy for the
model, it is not known {\em a priori} whether the nucleation of a
``supercritical" cluster will also be Poisson. To this end, in the
SC regime, we constructed cumulative probability distributions for
the lifetime of the resident allele $P_{\rm not}(t)$, i.e., the
probability that the global density of the resident has not
crossed below $\rho_{1}^{*}/2$ by time $t$.
We found that these distributions
are indeed exponentials (indicating that the nucleation of a
successful invading cluster is a Poisson process):
$P_{\rm not}(t)=1$ for $t$$\leq$$t_g$ and
$P_{\rm not}(t)=\exp[-(t-t_g)/\langle t_n\rangle]$ for $t$$>$$t_g$.
Here $\langle t_n\rangle$ is the average nucleation time and $t_g$ is
the close-to-deterministic growth time until the advantageous
mutation dominates half the system. We show results for a fixed
(sufficiently small) system size for three mutation rates in
Fig.~\ref{fig1}(a). From the slopes of the exponentials we
obtained the average nucleation times [Fig.~\ref{fig1}(b)], hence
the $\phi$-dependence of the nucleation rate per unit volume
$I(\phi)$. Since $\langle t_n\rangle=[L^{2}I(\phi)]^{-1}$, we have
$I(\phi)\sim\langle t_n\rangle^{-1}\sim\phi$. In the SC regime,
the invasive spread is inherently stochastic; it is initiated and
completed by the first randomly nucleated successful cluster of
the advantageous allele. For very low values of $\phi$, the
lifetime is dominated by the very large average nucleation times,
hence $\langle\tau\rangle = \langle t_n\rangle +t_g\approx \langle
t_n\rangle\sim\phi^{-1}$.

\begin{figure}[t]
\centering
\includegraphics[width=.48\textwidth,height=4.7truecm]{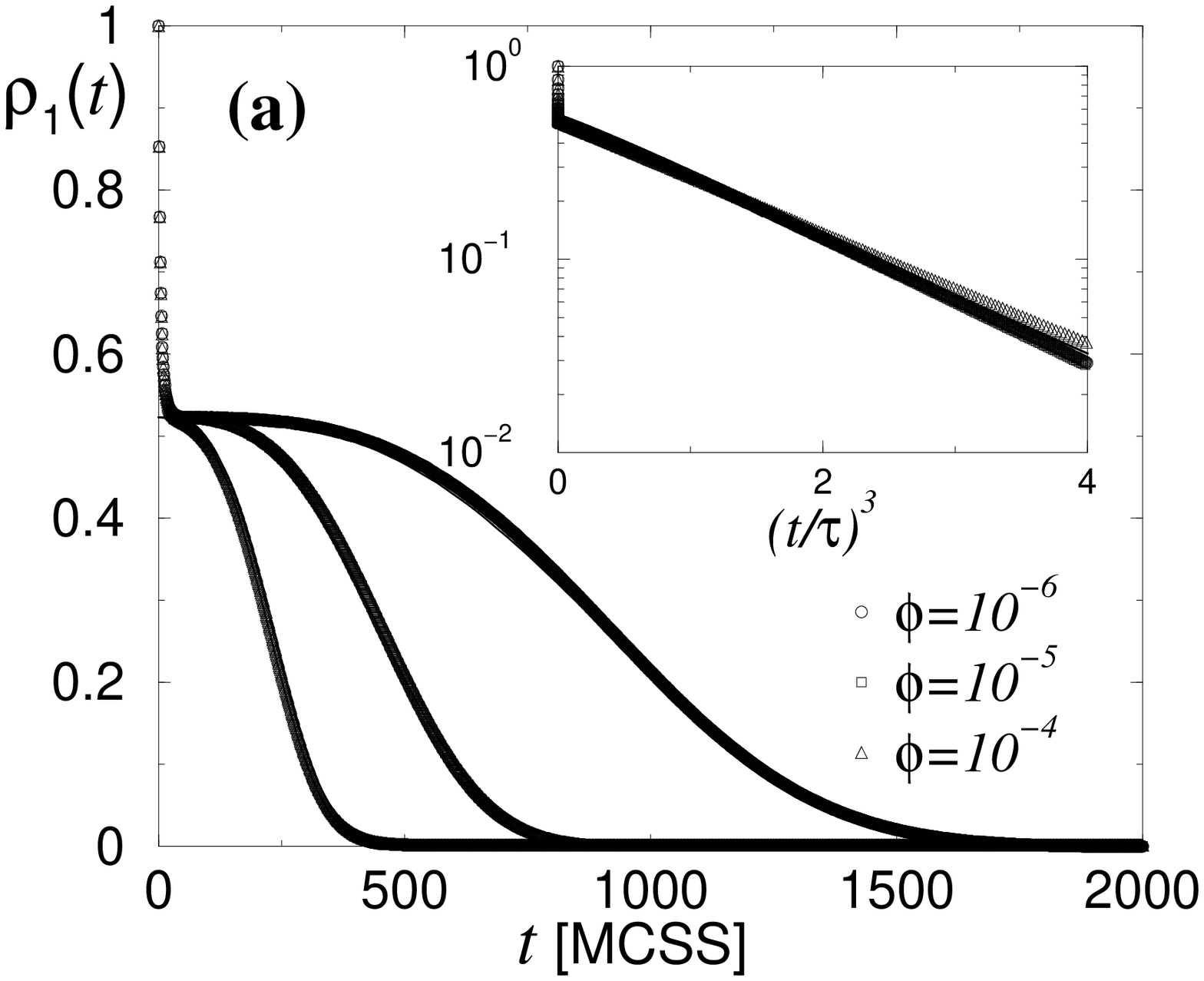}
\hspace*{0.2truecm}
\includegraphics[width=.48\textwidth,height=4.7truecm]{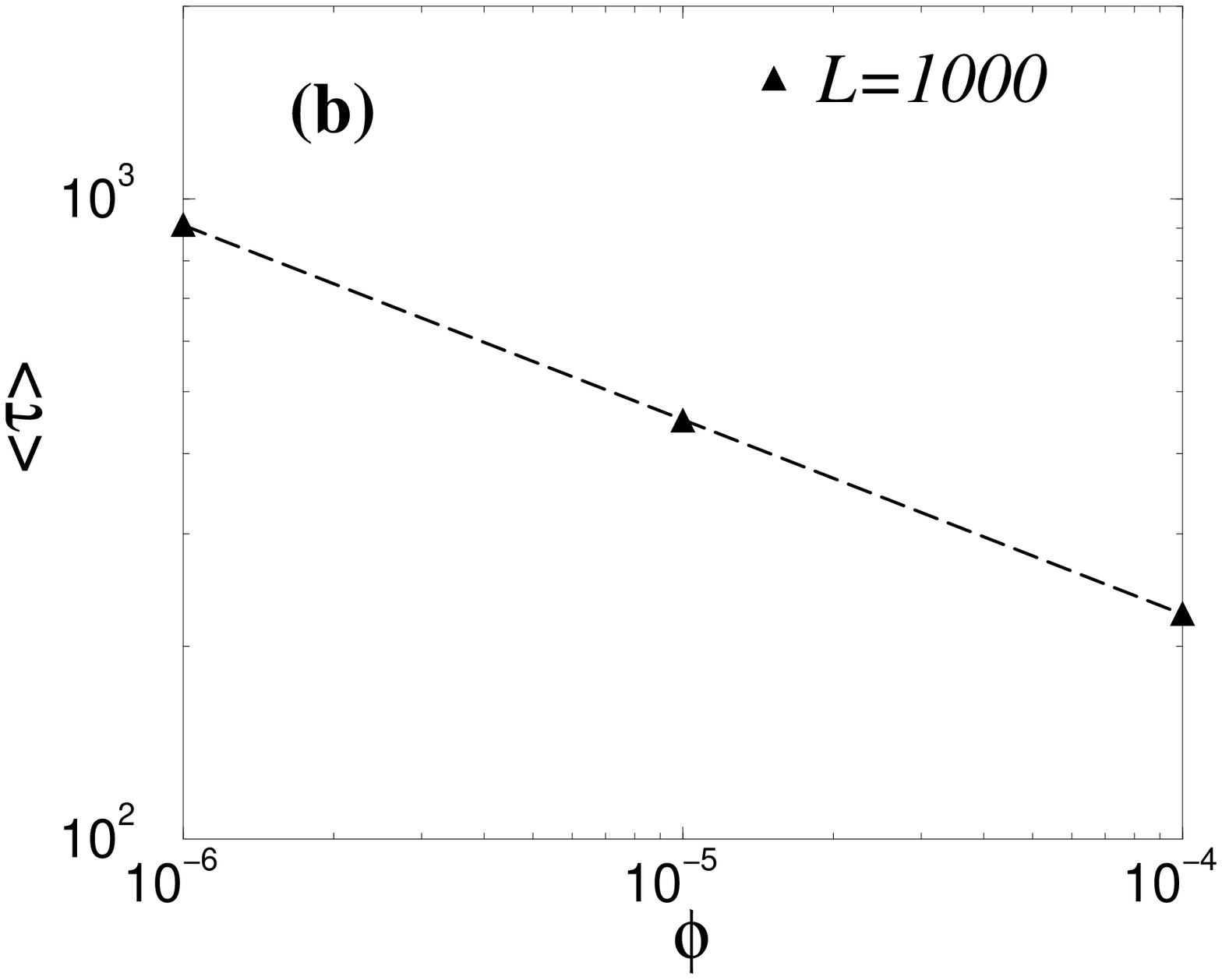}
\caption[]{
({\bf a}) Time-series of the global density of the resident allele $\rho_1(t)$ for $L$$=$$1,000$,
$\alpha_1$$=$$0.50$, $\alpha_2$$=$$0.70$, and $\mu$$=$$0.20$, for three different values of the
mutation rate $\phi$ (in increasing order from the top). The solid curves represent Avramis's law, Eq.~(\ref{avrami}).
The inset shows the  $\rho_1(t)$ vs. $t^3$ on log-linear scales.
({\bf b}) Average lifetime (in units of MCSS) vs. the mutation rate
on log-log scales for the same parameter values.
The straight dashed line is the best fit power-law indicating $\langle\tau\rangle\sim\phi^{-0.304}$.
}
\label{fig2}
\end{figure}

In the MC regime the invasion processes becomes self-averaging and
the global densities approach deterministic functions in the limit
of $L$$\to$$\infty$. At the same time, $\langle\tau\rangle$ approaches a
system-size independent limit. For large systems we applied the
KJMA theory \cite{KOLMOGOROV,JM,AVRAMI} (or Avrami's law) to predict the
density of the resident alelle,
\begin{equation}
\rho_{1}(t) \simeq \rho_{1}^{*} \, e^{-\ln(2)(t/\langle\tau\rangle)^3} \;.
\label{avrami}
\end{equation}
Our results in Fig~\ref{fig2}(a) show that it is, indeed, a very good approximation.
Assuming that the spreading velocity of the invading clusters is constant, KJMA
theory predicts that $\langle\tau\rangle\sim[I(\phi)]^{-1/3}\sim\phi^{-1/3}$ in the MC regime.
The measured exponent, $-0.304$, is not too far off [Fig.~\ref{fig2}(b)],
but it indicates that the assumption of a constant spreading velocity (possibly as a result
of the nontrivial surface properties of the clusters) may break down.

\section{Summary and Outlook}

We studied the spread of an advantageous mutant in a two-allele
population where rare mutations introduce the favored allele.
We found that nucleation theory, in particular Avrami's
law, describes this phenomenon very well. Systematic studies of the
critical cluster size and the cluster-size dependence of the
spreading velocity are under way. The
structure of the spreading clusters, in particular, the roughness
of their surface, is expected to play an important role in the latter.

%

%

\end{document}